\documentclass[aps,prl,showpacs,twocolumn,floats]{revtex4}
\usepackage{amsfonts}

\usepackage{graphicx,psfig}
\usepackage{dcolumn}
\usepackage{bm}

\input{tcilatex}

\begin{document}

\title{Field Dependence of the Electron Spin Relaxation in Quantum
Dots \vspace{-1mm} }
\author{Carlos Calero,$^1$ E. M. Chudnovsky,$^1$ and D. A. Garanin$^2$}

\affiliation{ \mbox{$^1$Department of Physics and Astronomy,
Lehman College, City University of New York,} \\ \mbox{250 Bedford
Park Boulevard West, Bronx, New York 10468-1589, U.S.A.} \\
\mbox{$^2$Institut f\"ur Physik, Johannes-Gutenberg-Universit\"at,
 D-55099 Mainz, Germany}}

\date{\today}
\begin{abstract}
Interaction of the electron spin with local elastic twists due to
transverse phonons has been studied. Universal dependence of the
spin relaxation rate on the strength and direction of the magnetic
field has been obtained in terms of the electron gyromagnetic
tensor and macroscopic elastic constants of the solid. The theory
contains no unknown parameters and it can be easily tested in
experiment. At high magnetic field it provides parameter-free
lower bound on the electron spin relaxation in quantum dots.
\end{abstract}
 \maketitle


Relaxation of the electron spin in solids is a fundamental problem that has
important applications. Among them are electron spin resonance and quantum
computing. In a semiconductor quantum dot the relaxation time for the
electron spin is determined by its interaction with phonons, nuclear spins,
impurities, etc. While impurities and nuclear spins can, in principle, be
eliminated, the interaction with phonons cannot. Thus, spin-phonon
interactions provide the most fundamental upper bound on the lifetime of
electron spin states. The existing methods of computing electron spin-phonon
rates in semiconductors rely upon phenomenological models of spin-orbit
interaction, see, e.g., Refs. %
\onlinecite{Dresselhaus,Rashba,Hasegawa,Roth,Khaetskii,Glavin,Rashba1,Efros,Loss,Tahan}%
. These models contain unknown constants that must be obtained
from experiment. Meantime, as has been noticed more than 50 years
ago by Elliot \cite{Elliot} (see also Ref.
\onlinecite{Serebrennikov}), the spin-orbit coupling in
semiconductors determines the difference of the electron g-factor
from the free electron value of $g_{0}=2.0023$. The question then
arises whether the effect of the spin-orbit coupling on
spin-phonon relaxation can be expressed via the difference between
the electron gyromagnetic tensor $g_{\alpha \beta }$ ($\alpha
,\beta =x,y,z$) and the vacuum tensor $g_{0}\delta _{\alpha \beta
}$. Since $g_{\alpha \beta }$ can be measured independently, this
would enable one to compare the computed relaxation rates with
experiment without any fitting parameters. In this Letter we show
that this, indeed, can be done under certain reasonable
simplifying assumptions.

Zeeman interaction of the electron with an external magnetic field, $\mathbf{%
B}$, is given by the Hamiltonian
\begin{equation}
\hat{H}_{Z}=-\mu _{B}\,g_{\alpha \beta }\,s_{\alpha }B_{\beta }\;,
\label{Zeeman}
\end{equation}
where $\mu _{B}$ is the Bohr magneton and
$\mathbf{s}=\mathbf{\sigma }/2$ is the dimensionless electron spin
with ${\sigma }_{\alpha }$ being Pauli matrices. One can choose
the axes of the coordinate system along the principal axes of the
tensor $g_{\alpha \beta }$. Then $g_{\alpha \beta }$ is diagonal,
\begin{equation}
g_{\alpha \beta }=g_{\alpha }\delta _{\alpha \beta },  \label{gdiag}
\end{equation}
represented by three numbers, $g_{x}$, $g_{y}$, and $g_{z}$ that can be
directly measured. Perturbation of Eq.\ (\ref{Zeeman}) by phonons has been
studied in the past \cite{Hasegawa,Roth,Glavin} by writing all terms of the
expansion of $g_{\alpha \beta }$ on the strain tensor, $u_{\alpha \beta }$,
permitted by symmetry. This gives spin-phonon interaction of the form $%
A_{\alpha \beta \gamma \delta }u_{\alpha \beta }{\sigma }_{\gamma }B_{\delta
}$ with unknown coefficients $A_{\alpha \beta \gamma \delta }$. To avoid
this uncertainty we limit our consideration to local rotations generated by
transverse phonons. The argument for doing this is three-fold. Firstly, the
rate of the transition accompanied by the emission or absorption of a phonon
is inversely proportional to the fifth power of the sound velocity \cite
{Abragam}. Since the velocity of the transverse sound is always smaller than
the velocity of the longitudinal sound \cite{LL-elasticity}, the transverse
phonons must dominate the transitions. Secondly, for a dot that is
sufficiently rigid to permit only tiny local rotations as a whole under an
arbitrary elastic deformation, the emission or absorbtion of a quantum of
the elastic twist will be the only spin-phonon relaxation mode. Finally, we
notice that interaction of the electron spin with a local elastic twist
generated by a transverse phonon does not contain any unknown constants.
Consequently, it gives parameter-free lower bound on the electron spin
relaxation rate.

The angle of the local rotation of the crystal lattice in the presence of
the deformation, $\mathbf{u}(\mathbf{r})$, is given by \cite{LL-elasticity}
\begin{equation}
\delta \mathbf{\phi }=\frac{1}{2}{\nabla }\times \mathbf{u},  \label{phi}
\end{equation}
and the local angular velocity is $\delta \dot{\mathbf{\phi }}$. The
analysis of the effect of the rotation on the electron spin can be done in
the coordinate frame that is rigidly coupled to the crystal lattice. In that
coordinate frame the effect of the rotation is two-fold. Firstly, it results
in the opposite rotation of the external magnetic field felt by the spin.
The corresponding perturbation of the magnetic field is given by $\delta
\mathbf{B}=\mathbf{B}\times \delta \mathbf{\phi }$. Secondly, the
Hamiltonian in the rotating frame acquires a kinematic term $-\hbar \mathbf{s%
}\cdot \delta \dot{\mathbf{\phi }}$. The spin-phonon interaction in the
lattice frame (marked by prime) is then given by
\begin{equation}
\hat{H}_{\mathrm{s-ph}}^{\prime }=-\hbar \,\mathbf{\Omega }^{\prime }\cdot
\mathbf{s},\qquad \Omega _{\alpha }^{\prime }=\delta \dot{\phi}_{\alpha
}+(\mu _{B}/\hbar )g_{\alpha \beta }\,\left[ \mathbf{B}\times \delta \mathbf{%
\phi }\right] _{\beta }.  \label{Hint}
\end{equation}
In these formulas $\delta \mathbf{\phi }$ should be understood as an
operator. Summation over repeated indices is implied. The canonical
quantization of phonons and Eq.\ (\ref{phi}) yield
\begin{eqnarray}
\mathbf{u} &=&\sqrt{\frac{\hbar }{2MN}}\sum_{\mathbf{k}\lambda }\frac{%
\mathbf{e}_{\mathbf{k}\lambda }e^{i\mathbf{k\cdot r}}}{\sqrt{\omega _{%
\mathbf{k}\lambda }}}\left( a_{\mathbf{k}\lambda }+a_{-\mathbf{k}\lambda
}^{\dagger }\right)   \label{uQuantized} \\
\delta \mathbf{\phi } &=&\frac{1}{2}\sqrt{\frac{\hbar }{2MN}}\sum_{\mathbf{k}%
\lambda }\frac{\left[ i\mathbf{k}\times \mathbf{e}_{\mathbf{k}\lambda }%
\right] e^{i\mathbf{k\cdot r}}}{\sqrt{\omega _{\mathbf{k}\lambda
}}}\left( a_{\mathbf{k}\lambda }+a_{-\mathbf{k}\lambda }^{\dagger
}\right)\,, \nonumber \\
\label{deltaphiPh}
\end{eqnarray}
where $M$ is the mass of the unit cell, $N$ is the number of cells in the
crystal, $\mathbf{e}_{\mathbf{k}\lambda }$ are unit polarization vectors, $%
\lambda =t_{1},t_{2},l$ denotes polarization, and $\omega _{k\lambda
}=v_{\lambda }k$ is the phonon frequency. The total Hamiltonian in the
lattice frame is
\begin{equation}
\hat{H}^{\prime }=\hat{H}_{0}+\hat{H}_{\mathrm{s-ph}}^{\prime }\,,\qquad
\hat{H}_{0}=\hat{H}_{Z}+\hat{H}_{\mathrm{ph}},  \label{H-total}
\end{equation}
where $\hat{H}_{Z}$ is Zeeman Hamiltonian of Eq. (\ref{Zeeman}) unperturbed
by phonons and $\hat{H}_{\mathrm{ph}}$ is Hamiltonian of free phonons.

Spin-phonon transitions occur between the eigenstates of $\hat{H}_{0}$.
These eigenstates are direct products of the spin and phonon states
\begin{equation}
\left| \Psi _{\pm }\right\rangle =\left| \psi _{\pm }\right\rangle \otimes
\left| \phi _{\pm }\right\rangle \,.  \label{eigenstates1}
\end{equation}
Here $\left| \psi _{\pm }\right\rangle $ are the eigenstates of $\hat{H}_{Z}$
with energies $E_{\pm }$ and $\left| \phi _{\pm }\right\rangle \ $ are the
eigenstates of $\hat{H}_{\mathrm{ph}}$ with energies $E_{\mathrm{ph}\pm }$,
satisfying
\begin{equation}
E_{+}+E_{\mathrm{ph},+}=E_{-}+E_{\mathrm{ph},-}\;.  \label{conservation}
\end{equation}
For $\hat{H}_{\mathrm{s-ph}}^{\prime }$ of Eq.\ (\ref{Hint}), which is
linear in phonon amplitudes, the states $\left| \phi _{\pm }\right\rangle $
differ by one emitted or absorbed phonon with a wave vector $\mathbf{k}$. We
will use the following designations
\begin{equation}
\left| \phi _{+}\right\rangle \equiv \left| n_{\mathbf{k}}\right\rangle
,\qquad \left| \phi _{-}\right\rangle \equiv \left| n_{\mathbf{k}%
}+1\right\rangle .  \label{phonon states}
\end{equation}

We need to compute the matrix element corresponding to the decay of the spin
$\left| \Psi _{+}\right\rangle \rightarrow \left| \Psi _{-}\right\rangle $.
With the help of Eq.\ (\ref{Hint}) we get:
\begin{equation}
\left\langle \Psi _{-}\right| \hat{H}_{\mathrm{s-ph}}^{\prime }\left| \Psi
_{+}\right\rangle =\mathbf{K\cdot }\,\left\langle \phi _{-}\right| \delta
\mathbf{\phi }\left| \phi _{+}\right\rangle ,  \label{matrix element}
\end{equation}
where components of vector $\mathbf{K}$ are given by
\begin{equation}
K_{\gamma }\equiv -\mu _{B}\left( g_{\alpha }-g_{\beta }\right) B_{\beta
}\epsilon _{\alpha \beta \gamma }\left\langle \psi _{-}\left| s_{\alpha
}\right| \psi _{+}\right\rangle ,  \label{KgammaLatt}
\end{equation}
and the principal components of the gyromagnetic tensor,
$g_{\alpha }$, are defined by Eq. (\ref{gdiag}). To obtain Eq.
(\ref{KgammaLatt}), we have used the relation
\begin{equation}
\delta \mathbf{\dot{\phi}}\,=\frac{i}{{\hbar
}}[\hat{H}_{\mathrm{ph}},\delta \mathbf{\phi }],  \label{phi-dot}
\end{equation}
to eliminate $\delta \mathbf{\dot{\phi}}$ from Eq.\ (\ref{Hint}),
and the energy conservation, Eq.\ (\ref{conservation}). Note that for the isotropic $%
g$-factor $K_{\gamma }=0$ and thus phonons do not couple to the spin.

As an independent test, one can consider the problem in the laboratory
frame. In the presence of the local rotation given by Eq.\ (\ref{phi}), the
gyromagnetic tensor in the laboratory frame becomes
\begin{equation}
g_{\alpha \beta }^{(\mathrm{ph})}={\Bbb{R}}_{\alpha \alpha ^{\prime }}{\Bbb{R%
}}_{\beta \beta ^{\prime }}g_{\alpha ^{\prime }\beta ^{\prime }},
\label{gtransf}
\end{equation}
where ${\Bbb{R}}_{\alpha \beta }$ is a $3\times 3$ rotation
matrix. For small $\delta \mathbf{\phi }$, one has
\begin{equation}
{\Bbb{R}}_{\alpha \beta }=\delta _{\alpha \beta }-\epsilon _{\alpha \beta
\gamma }\delta \phi _{\gamma }.  \label{rotation-matrix}
\end{equation}
Substituting Eq.\ (\ref{gtransf}) into Eq.\ (\ref{Zeeman}) and using the
orthogonality of the rotation matrix, ${\Bbb{R}}_{\alpha \beta }={\Bbb{R}}%
_{\beta \alpha }^{-1},$ we get for the Zeeman Hamiltonian in the presence of
phonons
\begin{equation}
\hat{H}_{Z}^{(\mathrm{ph})}=-\mu _{B}g_{\alpha ^{\prime }\beta ^{\prime
}}\left( {\Bbb{R}}_{\alpha ^{\prime }\alpha }^{-1}s_{\alpha }\right) \left( {%
\Bbb{R}}_{\beta ^{\prime }\beta }^{-1}B_{\beta }\right) .  \label{HZphDef}
\end{equation}
In the linear order in $\delta \mathbf{\phi }$ with the help of Eq. (\ref
{rotation-matrix}) one obtains the full Hamiltonian in the laboratory frame
\begin{equation}
\hat{H}=\hat{H}_{0}+\hat{H}_{\mathrm{s-ph}}\,,\qquad \hat{H}_{0}=\hat{H}_{Z}+%
\hat{H}_{\mathrm{ph}},  \label{H-total-lab}
\end{equation}
where
\begin{equation}
\hat{H}_{\mathrm{s-ph}}=-\hbar \mathbf{\Omega \cdot s,}  \label{HsphOmega}
\end{equation}
and
\begin{equation}
\Omega _{\alpha }=(\mu _{B}/\hbar )\left( g_{\alpha }-g_{\beta }\right)
B_{\beta }\epsilon _{\alpha \beta \gamma }\delta \phi _{\gamma }.
\label{Omegaalpha}
\end{equation}
One can see that the spin-phonon matrix element in the laboratory frame, $%
\left\langle \Psi _{-}\right| \hat{H}_{\mathrm{s-ph}}\left| \Psi
_{+}\right\rangle ,$ is the same as that in the lattice frame, Eqs. (\ref
{matrix element}) and (\ref{KgammaLatt}).

To obtain the relaxation rate one can use the Fermi golden rule. With the
help of Eq.\thinspace (\ref{deltaphiPh}) the transition matrix element of
Eq.\thinspace (\ref{matrix element}) can be expressed as
\begin{eqnarray}
&&\left\langle \Psi _{-}\right| \hat{H}_{\mathrm{s-ph}}\left| \Psi
_{+}\right\rangle   \nonumber  \label{matrix element2} \\
&&=\frac{\hbar }{\sqrt{N}}\sum_{\mathbf{k}\lambda }V_{\mathbf{k}\lambda
}\left\langle n_{\mathbf{k^{\prime }}}+1\right| a_{\mathbf{k}\lambda }+a_{-%
\mathbf{k}\lambda }^{\dagger }\left| n_{\mathbf{k^{\prime }}}\right\rangle ,
\end{eqnarray}
where
\begin{equation}
V_{\mathbf{k}\lambda }\equiv \frac{e^{i\mathbf{k}\mathbf{r}}}{\sqrt{8M\hbar
\omega _{\mathbf{k}\lambda }}}\,\mathbf{K}\cdot \lbrack \mathbf{k}\times
\mathbf{e}_{\mathbf{k}\lambda }].  \label{VDef}
\end{equation}
Note that only the transverse phonons contribute to the relaxation process.
The decay rate $W_{-+}$ of the upper spin state into the lower state,
accompanied by the emission of a phonon, and the rate $W{+-}$ of the inverse
process are given by
\begin{equation}
\left(
\begin{array}{c}
W_{-+} \\
W_{+-}
\end{array}
\right) =W_{0}\left(
\begin{array}{c}
n_{\omega _{0}}+1 \\
n_{\omega _{0}}
\end{array}
\right) ,  \label{rate}
\end{equation}
where $n_{\omega _{0}}=(e^{\hbar \omega _{0}/\left( k_{B}T\right) }-1)^{-1}$
is the phonon occupation number at equilibrium,
\begin{equation}
\hbar \omega _{0}\equiv E_{+}-E_{-}=\mu _{B}\left( \sum_{\gamma }g_{\gamma
}^{2}B_{\gamma }^{2}\right) ^{1/2}\,  \label{omega0final}
\end{equation}
is the distance between the two spin levels, and
\begin{equation}
W_{0}=\frac{1}{N}\sum_{\mathbf{k}\lambda }\left| V_{\mathbf{k}\lambda
}\right| ^{2}2\pi \delta (\omega _{\mathbf{k}\lambda }-\omega _{0}).
\label{W0}
\end{equation}
The balance equation for normalized populations of the upper spin state $%
n_{+}$ and the lower spin state $n_{-}$ (satisfying $n_{+}+n_{-}=1$ ) is
\begin{equation}
\dot{n}_{+}=-W_{-+}n_{+}+W_{+-}n_{-}=-\Gamma \,n_{+}+W_{+-}  \label{balance}
\end{equation}
where the relaxation rate is given by
\begin{equation}
\Gamma =W_{+-}+W_{-+}=W_{0}\coth \left(\frac{\hbar \omega
_{0}}{2k_{B}T}\right). \label{relaxation rate}
\end{equation}
Using Eq. (\ref{VDef}) and replacing $\left( 1/N\right) \sum_{\mathbf{k}%
}\ldots$ by $v_{0}\int d^{3}k/(2\pi )^{3}\ldots $ in Eq.\
(\ref{W0}), $v_{0}$ being unit-cell volume, one obtains
\begin{equation}
W_{0}=\frac{1}{12\pi \hbar }\frac{\left| \mathbf{K}\right| ^{2}}{Mv_{t}^{2}}%
\frac{\omega _{0}^{3}}{\omega _{D}^{3}}=\frac{1}{12\pi \hbar }\frac{\left|
\mathbf{K}\right| ^{2}\omega _{0}^{3}}{\rho v_{t}^{5}}.  \label{W0Res}
\end{equation}
Here $v_{t}$ is the velocity of the transverse sound, $\rho $ is
the mass density, $\omega _{D}\equiv v_{t}v_{0}^{1/3}$ is the
Debye frequency for the transverse phonons, and $\left|
\mathbf{K}\right| ^{2}\equiv \sum K_{\gamma }^{\ast }K_{\gamma }$.
With the help of Eq. (\ref{KgammaLatt}) we get
\begin{equation}
\left| \mathbf{K}\right| ^{2}=\mu _{B}^{2}\sum_{\alpha \beta
}\left( g_{\alpha }-g_{\beta }\right) ^{2}\left( B_{\beta
}^{2}T_{\alpha \alpha }+B_{\beta }B_{\alpha }T_{\alpha \beta
}\right) ,  \label{K2viaTalbet}
\end{equation}
where
\begin{equation}
T_{\alpha \beta }\equiv \frac{1}{2}\left( \left\langle \psi
_{-}\left| s_{\alpha }\right| \psi _{+}\right\rangle ^{\ast
}\left\langle \psi _{-}\left| s_{\beta }\right| \psi
_{+}\right\rangle +\text{c. c.}\right) . \label{TalbetDef}
\end{equation}

The eigenstates of $\hat{H}_{Z}$ are
\begin{eqnarray}
\left| \psi _{+}\right\rangle  &=&-\sin (\theta /2)e^{-i\varphi
/2}\left| +\right\rangle +\cos (\theta /2)e^{i\varphi /2}\left|
-\right\rangle   \nonumber
\\
\left| \psi _{-}\right\rangle  &=&\cos (\theta /2)e^{-i\varphi
/2}\left| +\right\rangle +\sin (\theta /2)e^{i\varphi /2}\left|
-\right\rangle, \label{eigenstates2}
\end{eqnarray}
where $\left| +\right\rangle ,\left| -\right\rangle $ are the
eigenstates of the operator $s_{z}$,
\begin{equation}
s_{z}\left| \pm \right\rangle =\pm \frac{1}{2}\left| \pm
\right\rangle , \label{z-eigenstates}
\end{equation}
and the spherical angles $\theta$ and $\varphi$ are defined
through
\begin{eqnarray}
\mathbf{b} &=& g_{x}H_{x}\mathbf{e}_{x}+g_{y}H_{y}\mathbf{e}%
_{y}+g_{z}H_{z}\mathbf{e}_{z}\,  \nonumber \\
&=&\left| \mathbf{b}\right| \left( \sin \theta \cos \varphi \,\mathbf{e}%
_{x}+\sin \theta \sin \varphi \,\mathbf{e}_{y}+\cos \theta \,\mathbf{e}%
_{z}\right). \nonumber \\  \label{pseudo-field}
\end{eqnarray}
This gives
\begin{eqnarray}\label{spin matrix elements}
\left\langle \psi_{-}\right| s_x \left| \psi_{+} \right\rangle
& = & \frac{1}{2}\left( i\sin \varphi +\cos \theta \cos \varphi \right)\nonumber\\
\left\langle \psi_{-}\right| s_y \left| \psi_{+}
\right\rangle & = & \frac{%
1}{2}\left( -i\cos \varphi +\cos \theta \sin \varphi \right) \nonumber\\
\left\langle \psi_{-}\right| s_z \left| \psi_{+}
\right\rangle & = & -\frac{1}{2}\sin \theta \,. \nonumber\\
\end{eqnarray}
Direct calculation then yields
\begin{equation}
T_{\alpha \beta }=\frac{1}{4}\left( \delta _{\alpha \beta
}-\frac{g_{\alpha }B_{\alpha }\,g_{\beta }B_{\beta }}{\sum_{\gamma
}\left( g_{\gamma }B_{\gamma }\right) ^{2}}\right) .
\label{TalbetGeneralFinal}
\end{equation}
Thus one obtains
\begin{eqnarray}
\left| \mathbf{K}\right| ^{2} &=&\frac{\mu _{B}^{2}}{8}\sum_{\alpha \beta
=x,y,z}\left( g_{\alpha }-g_{\beta }\right) ^{2}  \nonumber \\
&&\quad \times \left[ B_{\alpha }^{2}+B_{\beta }^{2}-\frac{\left( g_{\alpha
}+g_{\beta }\right) ^{2}B_{\alpha }^{2}B_{\beta }^{2}}{\sum_{\gamma }\left(
g_{\gamma }B_{\gamma }\right) ^{2}}\right] .  \label{K2Final}
\end{eqnarray}

With the help of Eq. (\ref{omega0final}), Eq. (\ref{relaxation
rate}) now can be written in the final form
\begin{equation}
\Gamma =\frac{1}{3\pi \hbar }\,\frac{\left( \mu _{B}B\right) ^{5}}{%
Mv_{t}^{2}\left( \hbar \omega _{D}\right) ^{3}}\,F_{T}\left(
\mathbf{n}\right) =\frac{\hbar}{3\pi \rho }\left(\frac{\mu
_{B}B}{\hbar v_t}\right)^{5}F_{T}\left( \mathbf{n}\right) ,
\label{rate2}
\end{equation}
where $\mathbf{n\equiv B/}B$ and
\begin{eqnarray}
F_{T}\left( \mathbf{n}\right) & = & \frac{1}{32}\left(
\sum_{\gamma }g_{\gamma }^{2}n_{\gamma
}^{2}\right) ^{3/2}\coth \left[ \frac{\mu _{B}B}{2k_{B}T}%
\left( \sum_{\gamma }g_{\gamma }^{2}n_{\gamma }^{2}\right) ^{1/2}\right]
\nonumber \\
&&\times \sum_{\alpha \beta }\left( g_{\alpha }-g_{\beta }\right)
^{2}\left[ n_{\alpha }^{2}+n_{\beta }^{2}-\frac{\left( g_{\alpha
}+g_{\beta }\right) ^{2}n_{\alpha }^{2}n_{\beta
}^{2}}{\sum_{\gamma }\left( g_{\gamma }n_{\gamma
}\right) ^{2}}\right] .  \nonumber \\
&&  \label{FTDef}
\end{eqnarray}
Here $\alpha ,\beta ,\gamma $ run over $x,y,z$. \ It is apparent from Eq.\ (%
\ref{FTDef}) that the relaxation mechanism studied in this Letter requires
anisotropy of the gyromagnetic tensor. When the field is directed along the $%
z$-axis, Eq.\ (\ref{FTDef}) simplifies to
\begin{equation}
F_{T}(\mathbf{e}_{z})=\frac{g_{z}^{3}}{16}[(g_{z}-g_{x})^{2}+(g_{z}-g_{y})^{2}]\coth
\left( \frac{g_{z}\mu _{B}B}{2k_{B}T}\right) \,.  \label{Fz}
\end{equation}

For the theory to be valid, $\omega_0$ of Eq.\ (\ref{omega0final})
should not exceed $\omega_D$, otherwise there will be no acoustic
phonons responsible for the discussed spin-phonon relaxation
mechanism. If $g_{\alpha}$ are of order unity, this is equivalent
to the condition that the factor $(\mu_BB/\hbar v_t)$ in Eq.\
(\ref{rate2}), that has dimensionality of the wave vector, is less
than the Debye wave vector, $k_D = \omega_D/v_t$, for transverse
phonons. This condition is almost always satisfied in the
experimentally accessible field range. At $\hbar \omega _{0}\gg
k_{B}T$ the $\coth $ factor in equations (\ref {FTDef}) and
(\ref{Fz}) tends to one. In this case ${\Gamma }\propto B^{5}$
while $F_{T}$ depends only on the direction of the field with
respect to the principal axes of $g_{\alpha \beta }$. In the
opposite limit of $k_{B}T\gg \hbar \omega _{0}$, the relaxation
rate is proportional to $B^{4}T$ while its dependence on the
direction of the field is given by the factor
\begin{equation}
\sum_{\alpha \beta }\left( g_{\alpha }-g_{\beta }\right)
^{2}\left[ n_{\alpha }^{2}+n_{\beta }^{2}-\frac{\left( g_{\alpha
}+g_{\beta }\right) ^{2}n_{\alpha }^{2}n_{\beta
}^{2}}{\sum_{\gamma }\left( g_{\gamma }n_{\gamma }\right)
^{2}}\right] \;. \label{direction-factor}
\end{equation}

A nice property of the spin relaxation mechanism studied above is its
universal dependence on the strength and the direction of the magnetic
field. Due to $B^{5}$ in Eq.\ (\ref{rate2}) this mechanism can dominate
electron spin relaxation at high fields. For, e.g., $\rho \sim 5\,$g/cm$^{3}$
and $v_{t}\sim 2\times 10^{5}$cm/s, it gives $\Gamma \sim 3 \times 10^{4}$s$%
^{-1}F_{T}(\mathbf{n})$. The dependence of the rate on the direction of the
field, $F_{T}(\mathbf{n})$, is entirely determined by the difference between
principal values of the tensor $g_{\alpha \beta}$. Highly anisotropic $%
g_{\alpha \beta}$ has been theoretically predicted in
two-dimensional systems \cite{Winkler} and experimentally detected
in GaAs quantum wells \cite{Salis}. In the mK temperature range,
the spin relaxation times of order $100\,\mu $s have been observed
\cite{Hanson} in GaAs electron quantum dots in the field of order
$10$T. Note that equations (\ref{rate2})-(\ref {direction-factor})
allow a detailed comparison between theory and experiment for the
proposed mechanism of relaxation, which, thus, can be easily
confirmed or ruled out for a particular quantum dot.

In conclusion, we have studied electron spin relaxation in quantum dots due
to local rotations generated by transverse phonons. This is unavoidable
relaxation channel that occurs when the electron gyromagnetic tensor, $%
g_{\alpha \beta} $, is anisotropic. It can dominate spin
relaxation at high magnetic fields. The advantage of our theory is
that it expresses the effect of unknown spin-orbit interactions on
electron spin relaxation in terms of the tensor $ g_{\alpha \beta}
$ alone. The corresponding relaxation rate has universal
dependence on the strength and direction of the field with respect
to the principal axes of $g_{\alpha \beta}$. The important feature
of the proposed mechanism is that it does not involve any unknown
constants of the quantum dot and is entirely determined by the
three principal values of $g_{\alpha \beta}$, which can be
independently measured. This allows simple experimental test of
the proposed theory.

This work has been supported by the NSF Grant No. EIA-0310517.



\end{document}